\documentclass[twocolumn]{jpsj2}

\title{%
Spin Injection and Detection in a Mesoscopic Superconductor
at Low Temperatures
}

\author{%
Yositake {\sc Takane}
}

\inst{%
Department of Quantum Matter, Graduate School of Advanced Sciences of Matter,
Hiroshima University, Higashi-Hiroshima, Hiroshima 739-8530, Japan
}

\recdate{ \hspace{50mm} }

\abst{%
We theoretically study nonequilibrium spin transport in a superconducting wire
connected by tunnel junctions to two ferromagnetic metal wires, each of which
serves as an injector or detector of spin-polarized electron current.
We present a set of Boltzmann equations to determine nonequilibrium
quasiparticle distributions in this system,
and obtain an analytical expression for the nonlocal spin signal
in the case of small injection current.
It is shown that the quasiparticle distribution in the ferromagnetic metal
for detection strongly affects the magnitude of the spin signal.
At low temperatures, since nonequilibrium quasiparticles created by
the tunneling from the superconductor dominate thermally excited ones,
the spin signal becomes independent of temperature.
This explains the convergence of the spin signal with decreasing temperature
observed in a recent experiment by Poli \textit{et al}.
}

\kword{%
spin accumulation, spin imbalance, energy imbalance, spin signal}

\begin{document}

\maketitle

\section{Introduction}

Experimental studies on spin injection and detection in a normal metal
have attracted considerable attention recently in the field of
spintronics.~\cite{rf:johnson1,rf:jedema1,rf:jedema2,rf:kimura,rf:takahashi}
More than two decades ago, Johnson and Silsbee~\cite{rf:johnson1} performed
the first experiment on this subject by using a large normal metal sample
with two electrodes made of a ferromagnetic metal,
where each electrode serves as a spin injector or detector.
Spin-polarized electrons created near the injector diffuse in the normal metal,
and spin imbalance is transmitted to the detector
if spin-flip scattering does not suppress it.
They found an evidence of spin imbalance
by measuring an open-circuit voltage induced at the detector.
Several experiments using devices in the mesoscopic regime have been
reported to date.~\cite{rf:jedema1,rf:jedema2,rf:kimura}
The most popular device consists of a thin normal metal wire connected to
a few ferromagnetic metal wires.
In this system, we supply injection current $I_{\rm inj}$
with spin polarization $P_{\rm spin}$ into
the normal metal from one of ferromagnetic metals and measure an open-circuit
voltage between another ferromagnetic metal and the normal metal.
Let $V_{\rm p}$ ($V_{\rm ap}$) be the open-circuit voltage when
the magnetizations of the two ferromagnetic metals are parallel (antiparallel).
We are interested in the nonlocal spin signal defined by
\begin{align}
     \label{eq:def-Rs}
  R_{\rm spin} = \frac{V_{\rm p}-V_{\rm ap}}{I_{\rm inj}} ,
\end{align}
which crucially depends on the spin diffusion length $\lambda_{\rm sf}$
and the distance $d$ between the injection and detection points.
In the case where the normal metal and the two ferromagnetic metals are
connected by tunnel junctions, the spin signal is given
by~\cite{rf:jedema2,rf:takahashi}
\begin{align}
  R_{\rm spin} = P_{\rm spin}^{2} R_{\rm N}
                 {\rm e}^{-\frac{d}{\lambda_{\rm sf}}} ,
\end{align}
where $R_{\rm N} \equiv \rho_{\rm N}\lambda_{\rm sf}/A_{\rm N}$ with
$\rho_{\rm N}$ and $A_{\rm N}$ being the resistivity and
the cross-sectional area of the normal metal, respectively.

Spin injection and detection in a superconductor is also attracted considerable
attention.~\cite{rf:takahashi,rf:johnson2,
rf:gu,rf:shin,rf:miura,rf:urech,rf:poli}
Our primary interest focuses on how the spin signal is modified
by the transition to the superconducting state.
Takahashi and Maekawa~\cite{rf:takahashi} studied
this problem and predicted that
\begin{align}
     \label{eq:TM}
  R_{\rm spin} = \frac{1}{2f_{0}(\Delta)}P_{\rm spin}^{2} R_{\rm N}
                 {\rm e}^{-\frac{d}{\lambda_{\rm sf}}} ,
\end{align}
where $f_{0}(\Delta) = 1/({\rm exp}(\Delta/T)+1)$
with the superconducting energy gap $\Delta$ and temperature $T$.
This indicates that $R_{\rm spin}$ exponentially increases with decreasing $T$.
They claimed that this modification is caused by the increase of
spin resistivity due to the opening of the energy gap $\Delta$.
The increase of $R_{\rm spin}$ with decreasing $T$ has been successfully
observed in the recent experiment by Poli \textit{et al}.~\cite{rf:poli}
However, there remain a few points to be clarified.
We focus on the following two points.
Firstly, Takahashi and Maekawa implicitly assume in their derivation
of eq.~(\ref{eq:TM}) that spin imbalance in a superconductor can be described
by a shift of spin-dependent chemical potential.
This assumption cannot be justified at low temperatures,
where energy relaxation due to phonon scattering is not strong.
Secondly, Poli \textit{et al}. observed convergence of $R_{\rm spin}$
with decreasing temperature.
This behavior cannot be explained by eq.~(\ref{eq:TM}).

In this paper, we theoretically study nonequilibrium spin transport
in a hybrid system consisting of
a superconducting wire and two ferromagnetic metal wires.
Each ferromagnetic metal is connected by a tunnel junction
to the superconductor,
and serves as an injector or detector of spin-polarized quasiparticles.
We present a set of Boltzmann equations
governing nonequilibrium quasiparticles in this system.
We focus on the case of small injection current at low temperatures,
and obtain not only the quasiparticle distribution in the superconducting wire
but also that in the ferromagnetic metal wire for detection.
On the basis of the resulting nonequilibrium distributions,
we derive an analytical expression for the nonlocal spin signal.
It is shown that although the spin signal originates from spin imbalance
transmitted to the detection junction,
its magnitude is not solely determined by the spin imbalance but is strongly
affected by the quasiparticle distribution in the ferromagnetic metal.
We observe that when $T$ is higher than a crossover temperature
$T_{\rm cross}$, the spin signal exponentially increases with decreasing $T$
reflecting the reduction of thermally excited quasiparticles
in the ferromagnetic metal.
At low temperatures below $T_{\rm cross}$, however,
the magnitude of the spin signal is determined by nonequilibrium
quasiparticles created by the tunneling from the superconductor
instead of thermally excited ones,
and the spin signal becomes independent of $T$.
This explains the convergence of the spin signal with decreasing $T$
observed by Poli \textit{et al}.~\cite{rf:poli}

In the next section, we present a set of Boltzmann equations to describe
nonequilibrium quasiparticle distributions in the hybrid system consisting of
a superconducting wire and two ferromagnetic metal wires.
In \S 3, we obtain nonequilibrium quasiparticle distributions in this system
by solving the set of Boltzmann equations, and derive an analytical expression
of the spin signal on the basis of the resulting quasiparticle distributions.
In \S 4, we compare our theoretical result with the recent experimental result.
We set $\hbar = k_{\rm B} = 1$ throughout this paper.

\section{Formulation}

Let us consider the hybrid system consisting of a superconducting wire
and two ferromagnetic metal wires (see Fig.~1).
\begin{figure}[hbtp]
\begin{center}
\includegraphics[height=6cm]{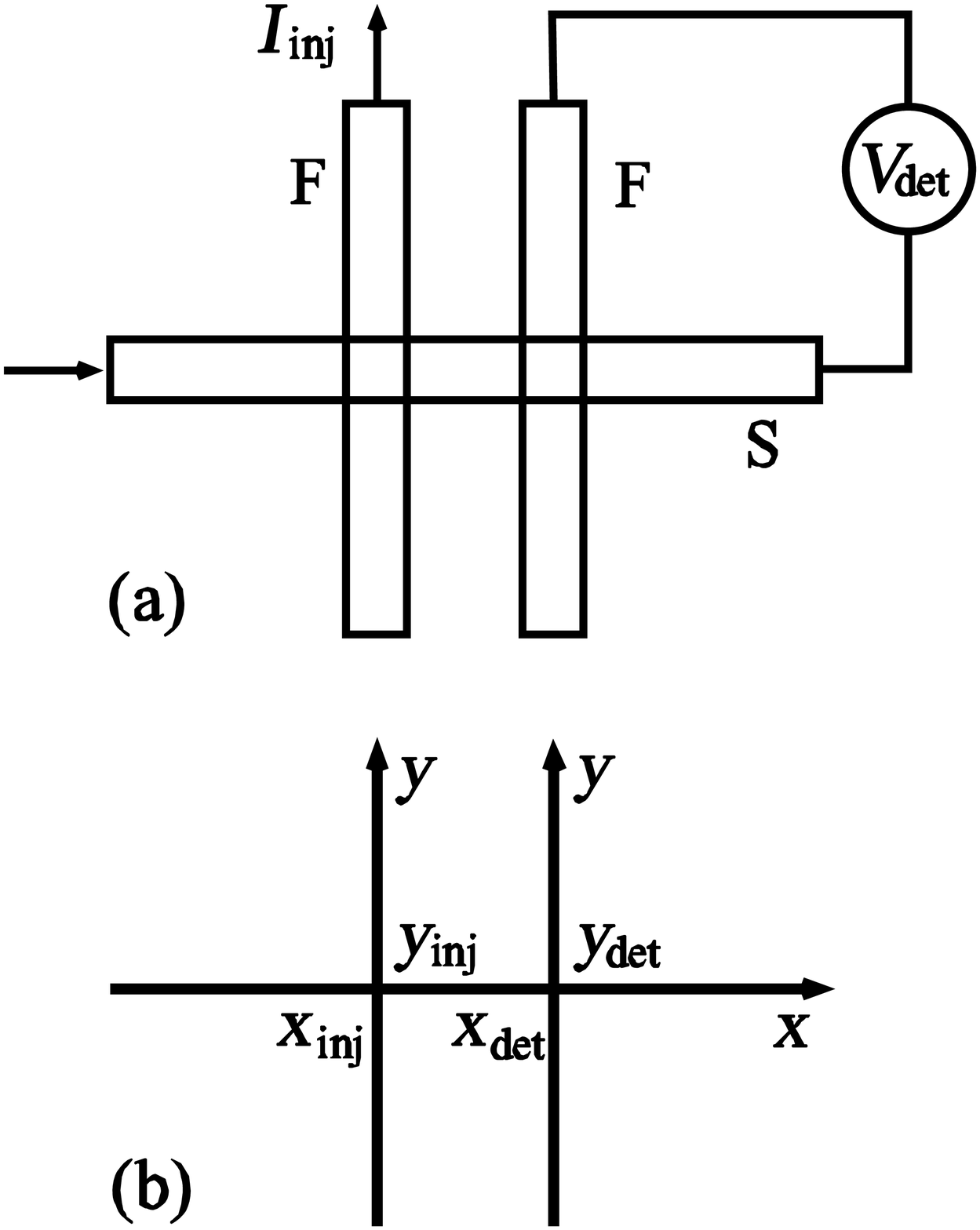}
\end{center}
\caption{(a) Schematic picture of the model system consisting of
a superconducting wire (S) and two ferromagnetic metal wires (F).
The left (right) ferromagnetic metal serves as an injector (detector)
of spin-polarized electron current.
(b) Spatial coordinates used in the text.
}
\end{figure}
We assume that the superconductor is connected by a tunnel junction
to each ferromagnetic metal.
The left and right junctions serve as spin injector and detector, respectively.
We adopt a simple one-dimensional model for this device
assuming that the superconductor and the ferromagnetic metals are very thin.
We introduce the $x$ axis in the superconductor on which the left and right
junctions are located at $x = x_{\rm inj}$ and $x = x_{\rm det}$, respectively,
and the $y$ axis in the left (right) ferromagnetic metal
on which the injection (detection) junction is located at $y = y_{\rm inj}$
($y_{\rm det}$).
We denote by $d$ the separation between the two junctions.
That is, $d \equiv x_{\rm det}-x_{\rm inj}$.
We inject spin-polarized current into the superconductor by applying a bias
voltage $V_{\rm inj}$ across the injection junction,
and measure an induced open-circuit voltage $V_{\rm det}$ across
the detection junction under the condition that net current flow vanishes
between the superconductor and the ferromagnetic metal for detection.
We simply assume that the spin polarization $P_{\rm spin}$ of the injection
current is proportional to the difference between the density of states
$N_{{\rm F}\uparrow}$ for up-spin electrons and
$N_{{\rm F}\downarrow}$ for down-spin electrons.
The spin polarization is expressed as
\begin{align}
 P_{\rm spin} = \frac{N_{{\rm F}\uparrow}-N_{{\rm F}\downarrow}}
                  {N_{{\rm F}\uparrow}+N_{{\rm F}\downarrow}} .
\end{align}
We assume that spin relaxation in the superconductor is caused by spin-flip
scattering due to spin-orbit interaction as well as magnetic impurities.

To present an expression for the tunneling current across each junction,
we consider nonequilibrium quasiparticle distributions in the superconductor
and the ferromagnetic metals.
We first introduce the quasiparticle distribution function
$g_{{\rm FL}\sigma}$ in the left ferromagnetic metal for injection,
where $\sigma = \uparrow, \downarrow$ is the spin variable.
We assume $g_{{\rm FL}\sigma}(y,\epsilon) = f_{0}(\epsilon-eV_{\rm inj})$
with the Fermi-Dirac distribution function $f_{0}(\epsilon)$.
Here and hereafter, we measure quasiparticle energy from the chemical potential
of the superconductor not only in the superconductor
but also in the ferromagnetic metals.
We next introduce the quasiparticle distribution function $g_{{\rm S}\sigma}$
in the superconductor.
In terms of four distribution functions
$f_{{\rm L}+}$, $f_{{\rm L}-}$, $f_{{\rm T}+}$ and $f_{{\rm T}-}$
for nonequilibrium quasiparticles,
we express it as~\cite{rf:schmid,rf:hu,rf:morten1,rf:morten2,rf:takane1}
\begin{align}
         \label{eq:f-up}
     g_{{\rm S}\uparrow}(x,\epsilon)
 & = f_{0}(\epsilon)
         + f_{{\rm L}+}(x,\epsilon) + f_{{\rm L}-}(x,\epsilon)
    \nonumber \\
 & \hspace{20mm}
         + f_{{\rm T}+}(x,\epsilon) + f_{{\rm T}-}(x,\epsilon) ,
            \\
         \label{eq:f-down}
     g_{{\rm S}\downarrow}(x,\epsilon)
 & = f_{0}(\epsilon)
         - f_{{\rm L}+}(x,\epsilon) + f_{{\rm L}-}(x,\epsilon)
    \nonumber \\
 & \hspace{20mm}
         + f_{{\rm T}+}(x,\epsilon) - f_{{\rm T}-}(x,\epsilon) .
\end{align}
The four distribution functions satisfy
\begin{align}
   f_{{\rm L,T}+}(x,-\epsilon) & = f_{{\rm L,T}+}(x,\epsilon) ,
           \\
   f_{{\rm L,T}-}(x,-\epsilon) & = - f_{{\rm L,T}-}(x,\epsilon) .
\end{align}
Note that $f_{{\rm L}+}$ describes spin imbalance, while $f_{{\rm T}+}$
describes charge imbalance.~\cite{rf:clarke,rf:tinkham}
The other two functions $f_{{\rm L}-}$ and $f_{{\rm T}-}$ describe
total energy imbalance and energy imbalance
between up-spin and down-spin quasiparticles, respectively.
Finally, we introduce the distribution function $g_{{\rm FR}\sigma}$
in the right ferromagnetic metal in which nonequilibrium quasiparticles appear
due to quasiparticle tunneling from the superconductor.
We express it as
\begin{align}
  g_{{\rm FR}\sigma}(y,\epsilon)
   = f_{0}(\epsilon-eV_{\rm det}) + f_{{\rm F}\sigma}(y,\epsilon) .
\end{align}
We hereafter assume that the magnitude of the energy gap $\Delta$ is unaffected
by spin injection everywhere in the superconductor.
This allows us to consider $f_{{\rm L}\pm}(x,\epsilon)$ and
$f_{{\rm T}\pm}(x,\epsilon)$ only for $|\epsilon| > \Delta$.
The nonequilibrium distribution functions
$f_{{\rm L}\pm}$, $f_{{\rm T}\pm}$ and $f_{{\rm F}\sigma}$
are governed by Boltzmann equations which we present below.

The tunneling current at the injection junction is given by
\begin{align}
       \label{eq:I_inj}
     I_{\rm inj}(V_{\rm inj})
   = \frac{\Delta}{eR_{\rm inj}} J_{1}(V_{\rm inj},T) ,
\end{align}
where $R_{\rm inj}$ is the tunnel resistance of the injection junction and
\begin{align}
  J_{1}(V,T)
          & = \frac{1}{\Delta}
              \int_{0}^{\infty}{\rm d}\epsilon N_{1}(\epsilon)
              \bigl(  f_{0} \left(\epsilon-eV\right)
                    - f_{0} \left(\epsilon+eV\right)
              \bigr) 
\end{align}
with $N_{1}$ being the normalized density of states in the superconductor,
given by $N_{1}(\epsilon) = |\epsilon|/\sqrt{\epsilon^{2}-\Delta^{2}}$
for $|\epsilon| > \Delta$ in the BCS limit.
In deriving eq.~(\ref{eq:I_inj}), we have ignored small contributions
arising from nonequilibrium quasiparticles in the superconductor.
The tunneling current between the superconductor and
the right ferromagnetic metal for detection is expressed as
\begin{align}
   I_{\rm det}(V_{\rm det})
 = I_{\rm q}(V_{\rm det}) + I_{\rm F}(V_{\rm det}) - I_{\rm S}(V_{\rm det}) ,
\end{align}
where $I_{\rm q}$ is the ordinary tunneling current arising from thermally
excited quasiparticles, while $I_{\rm F}$ represents the contribution
from nonequilibrium quasiparticles induced in the ferromagnetic metal.
The third term $I_{\rm S}$ represents the contribution from
spin and charge imbalances.
They are expressed as
\begin{align}
     I_{\rm q}(V_{\rm det})
 & = \frac{\Delta}{eR_{\rm det}} J_{1}(V_{\rm det},T) ,
   \\
        \label{eq:I_F}
     I_{\rm F}(V_{\rm det})
 & = \frac{1}{eR_{\rm det}} \int_{0}^{\infty} {\rm d}\epsilon \hspace{1mm}
     N_{1}(\epsilon)
    \nonumber \\
 & \hspace{-5mm}
     \times
     \biggl( \frac{1+P_{\rm spin}}{2}
              \left( f_{{\rm F}\uparrow}(y_{\rm det},\epsilon)
                     + f_{{\rm F}\uparrow}(y_{\rm det},-\epsilon) \right)
        \nonumber \\
 & \hspace{0mm}
            + \frac{1-P_{\rm spin}}{2}
              \left( f_{{\rm F}\downarrow}(y_{\rm det},\epsilon)
                     + f_{{\rm F}\downarrow}(y_{\rm det},-\epsilon) \right)
     \biggr) ,
   \\
        \label{eq:I_S}
     I_{\rm S}(V_{\rm det})
 & = \frac{2}{eR_{\rm det}} \int_{0}^{\infty} {\rm d}\epsilon \hspace{1mm}
     N_{1}(\epsilon)
    \nonumber \\
 & \hspace{5mm}
     \times
               \bigl(  P_{\rm spin} f_{{\rm L}+}(x_{\rm det}, \epsilon)
                     + f_{{\rm T}+}(x_{\rm det}, \epsilon)
               \bigr) ,
\end{align}
where $R_{\rm det}$ is the tunnel resistance of the detection junction.
In eq.~(\ref{eq:I_S}), the first term with $f_{{\rm L}+}$ represents
the contribution from spin imbalance and is the origin of the spin signal,
while the second term with $f_{{\rm T}+}$ represents
that from charge imbalance.
In deriving eqs.~(\ref{eq:I_F}) and (\ref{eq:I_S}), we have assumed
the parallel alignment of magnetizations.
The corresponding expressions for the antiparallel alignment is obtained
by reversing the sign of $P_{\rm spin}$.

To present Boltzmann equations for $f_{{\rm L}\pm}$ and
$f_{{\rm T}\pm}$, we introduce the Usadel equation~\cite{rf:usadel} for
the quasiclassical retarded Green's functions $g^{R}$ and $f^{R}$,
\begin{align}
     \label{eq:usadel}
 {\rm i} \epsilon f^{R}(\epsilon) + \Delta g^{R}(\epsilon)
    - \frac{1}{\tau_{\rm m}} g^{R}(\epsilon)f^{R}(\epsilon) = 0 ,
\end{align}
where $\tau_{\rm m}$ represents the magnetic impurity scattering time
and we have assumed that the superconductor is spatially homogeneous.
The spectral functions $N_{1}$, $N_{2}$, $R_{1}$ and $R_{2}$ are defined as
\begin{align}
   g^{R}(\epsilon) & = N_{1}(\epsilon) + {\rm i} R_{1}(\epsilon) ,
           \\
   f^{R}(\epsilon) & = N_{2}(\epsilon) + {\rm i} R_{2}(\epsilon) .
\end{align}
In terms of the spectral functions, the Boltzmann equations
are expressed as~\cite{rf:schmid,rf:hu,rf:morten1,rf:morten2,rf:takane1}
\begin{align}
        \label{eq:fL+}
  &  D_{\rm S} \left( N_{1}^{2}(\epsilon)-R_{2}^{2}(\epsilon) \right)
       \partial_{x}^{2} f_{{\rm L}+}(x,\epsilon)
    \nonumber \\
  & \hspace{10mm}
   - \frac{4}{3\tau_{\rm so}} \left( N_{1}^{2}(\epsilon)-R_{2}^{2}(\epsilon)
                             \right) f_{{\rm L}+}(x,\epsilon)
     \nonumber \\
  & \hspace{10mm}
  - \frac{4}{3\tau_{\rm m}} \left( N_{1}^{2}(\epsilon)+R_{2}^{2}(\epsilon)
                             \right) f_{{\rm L}+}(x,\epsilon)
     \nonumber \\
  & \hspace{10mm}
   + P_{{\rm L}+} (x,\epsilon) = 0 ,
               \\
        \label{eq:fL-}
  &  D_{\rm S} \left( N_{1}^{2}(\epsilon)-R_{2}^{2}(\epsilon) \right)
       \partial_{x}^{2} f_{{\rm L}-}(x,\epsilon)
   + P_{{\rm L}-} (x,\epsilon) = 0 ,
               \\
        \label{eq:fT+}
  &  D_{\rm S} \left( N_{1}^{2}(\epsilon)+N_{2}^{2}(\epsilon) \right)
       \partial_{x}^{2} f_{{\rm T}+}(x,\epsilon)
    \nonumber \\
  & \hspace{10mm}
   - \frac{1}{\tau_{\rm conv}(\epsilon)} f_{{\rm T}+}(x,\epsilon)
   + P_{{\rm T}+} (x,\epsilon) = 0 ,
               \\
        \label{eq:fT-}
  &  D_{\rm S} \left( N_{1}^{2}(\epsilon)+N_{2}^{2}(\epsilon) \right)
       \partial_{x}^{2} f_{{\rm T}-}(x,\epsilon)
     \nonumber \\
  & \hspace{10mm}
   - \frac{4}{3\tau_{\rm so}} \left( N_{1}^{2}(\epsilon)+N_{2}^{2}(\epsilon)
                              \right) f_{{\rm T}-}(x,\epsilon)
                 \nonumber \\
  & \hspace{10mm}
  - \frac{4}{3\tau_{\rm m}} \left( N_{1}^{2}(\epsilon)-N_{2}^{2}(\epsilon)
                             \right) f_{{\rm T}-}(x,\epsilon)
     \nonumber \\
  & \hspace{10mm}
   - \frac{1}{\tau_{\rm conv}(\epsilon)} f_{{\rm T}-}(x,\epsilon)
   + P_{{\rm T}-} (x,\epsilon) = 0 ,
\end{align}
where $D_{\rm S}$ is the diffusion constant, $\tau_{\rm so}$ and
$\tau_{\rm conv}$ are the spin-orbit scattering time and the charge imbalance
conversion time, respectively, and $P_{{\rm L}\pm}$ and $P_{{\rm T}\pm}$ are
the injection terms which represent quasiparticle tunneling between
the superconductor and the left ferromagnetic metal.
The injection terms are given as~\cite{rf:takane1,rf:takane2}
\begin{align}
       \label{eq:PL+}
   P_{{\rm L}+}(x,\epsilon)
& = \frac{\delta(x-x_{\rm inj})N_{1}(\epsilon)}
         {4e^{2}N_{\rm S}A_{\rm S}R_{\rm inj}}
     \nonumber \\
  & \hspace{-5mm}
    \times
             \Bigl[  P_{\rm spin}
                     \bigl(  f_{0} \left(\epsilon-eV_{\rm inj}\right)
                           - f_{0} \left(\epsilon+eV_{\rm inj}\right)
                     \bigr)
        \nonumber \\
& \hspace{0mm}
                 - 2 \bigl(  f_{{\rm L}+}(x_{\rm inj},\epsilon)
                           + P_{\rm spin} f_{{\rm T}+}(x_{\rm inj},\epsilon)
                     \bigr)
             \Bigr] ,
        \\
      \label{eq:PL-}
   P_{{\rm L}-}(x,\epsilon)
& = \frac{\delta(x-x_{\rm inj})N_{1}(\epsilon)}
         {4e^{2}N_{\rm S}A_{\rm S}R_{\rm inj}}
     \nonumber \\
  & \hspace{-5mm}
    \times
             \Bigl[  f_{0} \left(\epsilon+eV_{\rm inj}\right)
                   + f_{0} \left(\epsilon-eV_{\rm inj}\right)
                   - 2 f_{0} \left(\epsilon\right)
        \nonumber \\
& \hspace{0mm}
                - 2 \bigl(  f_{{\rm L}-}(x_{\rm inj},\epsilon)
                          + P_{\rm spin} f_{{\rm T}-}(x_{\rm inj},\epsilon)
                    \bigr)
             \Bigr] ,
        \\
      \label{eq:PT+}
   P_{{\rm T}+}(x,\epsilon)
& = \frac{\delta(x-x_{\rm inj})N_{1}(\epsilon)}
         {4e^{2}N_{\rm S}A_{\rm S}R_{\rm inj}}
     \nonumber \\
  & \hspace{-5mm}
    \times
             \Bigl[  f_{0} \left(\epsilon-eV_{\rm inj}\right)
                   - f_{0} \left(\epsilon+eV_{\rm inj}\right)
        \nonumber \\
& \hspace{0mm}
                - 2 \bigl(  P_{\rm spin} f_{{\rm L}+}(x_{\rm inj},\epsilon)
                          + f_{{\rm T}+}(x_{\rm inj},\epsilon)
                    \bigr)
             \Bigr] ,
        \\
      \label{eq:PT-}
   P_{{\rm T}-}(x,\epsilon)
& = \frac{\delta(x-x_{\rm inj})N_{1}(x,\epsilon)}
         {4e^{2}N_{\rm S}A_{\rm S}R_{\rm inj}}
     \nonumber \\
  & \hspace{-5mm}
    \times
             \Bigl[ P_{\rm spin}
                    \bigl(  f_{0} \left(\epsilon+eV_{\rm inj}\right)
                          + f_{0} \left(\epsilon-eV_{\rm inj}\right)
                          - 2 f_{0} \left(\epsilon\right)
                    \bigr)
        \nonumber \\
& \hspace{0mm}
                  - 2
                    \bigl(  P_{\rm spin} f_{{\rm L}-}(x_{\rm inj},\epsilon)
                          + f_{{\rm T}-}(x_{\rm inj},\epsilon)
                    \bigr)
             \Bigr] ,
\end{align}
where $N_{\rm S}$ and $A_{\rm S}$ are the density of states at the Fermi level
in the normal state and the cross-sectional area of the superconducting wire,
respectively.
We can ignore $f_{{\rm L}\pm}(x_{\rm inj},\epsilon)$ and
$f_{{\rm T}\pm}(x_{\rm inj},\epsilon)$ in these injection terms
when the injection current is small.
It should be noted that inelastic phonon scattering has been ignored in
the Boltzmann equations because its role is not relevant at low temperatures,
in which we are interested.
We have also ignored very small contributions to
$f_{{\rm L}\pm}$ and $f_{{\rm T}\pm}$
arising from the coupling with the right ferromagnetic metal for detection.

We turn to quasiparticle distributions in the right ferromagnetic metal
for detection.
We note that in obtaining $I_{\rm F}$, the spin-dependence of 
$f_{{\rm F}\sigma}$ is not important as long as the spin polarization is small.
This indicates that we need not consider complicated spin-dependent
dynamics of nonequilibrium quasiparticles.
We thus define
\begin{align}
  f_{{\rm F}+}(y,\epsilon)
 = \frac{  f_{{\rm F}\uparrow}(y,\epsilon)
         + f_{{\rm F}\downarrow}(y,\epsilon)}
        {2} ,
\end{align}
and approximate the expression of $I_{\rm F}$ as
\begin{align}
        \label{eq:I_F_mod}
     I_{\rm F}(V_{\rm det})
  & = \frac{1}{eR_{\rm det}} \int_{0}^{\infty} {\rm d}\epsilon \hspace{1mm}
     N_{1}(\epsilon)
     \nonumber \\
  & \hspace{5mm}
    \times
     \bigl( f_{{\rm F}+}(y_{\rm det},\epsilon)
             + f_{{\rm F}+}(y_{\rm det},-\epsilon)
     \bigr) .
\end{align}
We present an appropriate Boltzmann equation for $f_{{\rm F}+}$.
We ignore roles of spin-flip scattering
since the spin-dependence is not important for our argument.
However, we must consider the energy relaxation process
due to phonon scattering.
The reason for this is as follows.
Since quasiparticles in the ferromagnetic metal are induced by the tunneling
from the superconductor with the energy gap $\Delta$,
their excitation energy is of the order of $\Delta$ and
no quasiparticle is directly created in the subgap region.
Quasiparticles in such a nonequilibrium situation
inevitably experience the energy relaxation.
We thus assume that $f_{{\rm F}+}$ obeys
\begin{align}
       \label{eq:fF+}
     D_{\rm F} \partial_{y}^{2} f_{{\rm F}+}(y,\epsilon)
   - \frac{1}{\tau_{\rm e}(\epsilon-eV_{\rm det})} f_{{\rm F}+}(y,\epsilon)
   + P_{{\rm F}+} (y,\epsilon) = 0 ,
\end{align}
where $D_{\rm F}$ is the diffusion constant averaged over spin directions and
$\tau_{\rm e}$ is the energy relaxation time with $\epsilon-eV_{\rm det}$
being the quasiparticle energy measured
from the chemical potential of the ferromagnetic metal.
The source term $P_{{\rm F}+}$ describing quasiparticle tunneling
from the superconductor is given by
\begin{align}
     P_{{\rm F}+}(y,\epsilon)
 & = \frac{\delta(y-y_{\rm det})N_{1}(\epsilon)}
          {2e^{2}N_{\rm F}A_{\rm F}R_{\rm det}}
     \bigl(   f_{0}\left(\epsilon\right)
              - f_{0} \left(\epsilon-eV_{\rm det}\right)
     \nonumber \\
 & \hspace{0mm}
              - f_{{\rm F}+}(y_{\rm det},\epsilon)
              + f_{{\rm L}-}(x_{\rm det},\epsilon)
              + f_{{\rm T}+}(x_{\rm det},\epsilon)
     \bigr) ,
\end{align}
where $N_{\rm F} \equiv (N_{{\rm F}\uparrow}+N_{{\rm F}\downarrow})/2$
and $A_{\rm F}$ is the cross-sectional area of the ferromagnetic metal.
For the expression of $\tau_{\rm e}$, we adopt
\begin{align}
     \frac{1}{\tau_{\rm e}(\epsilon)}
  & = 2 \int_{-\infty}^{\infty} {\rm d}\epsilon'
     \sigma_{\rm F}(\epsilon, \epsilon')
     \nonumber \\
  & \hspace{5mm}
    \times
     \left( \coth \left(\frac{\epsilon'-\epsilon}{2T}\right)
             - \tanh \left(\frac{\epsilon'}{2T}\right)
     \right)
\end{align}
with
\begin{align}
        \label{eq:sigma_F}
     \sigma_{\rm F}(\epsilon, \epsilon')
   = \frac{\alpha_{\rm F}}{4} {\rm sign}(\epsilon'-\epsilon)
     \times (\epsilon'-\epsilon)^{2} ,
\end{align}
where $\alpha_{\rm F}$ characterizes
the strength of electron-phonon coupling.
For $|\epsilon| \gg T$, we approximately obtain
\begin{align}
     \frac{1}{\tau_{\rm e}(\epsilon)}
   = \frac{\alpha_{\rm F}}{3} |\epsilon|^{3} .
\end{align}

\section{Spin Signal}

In this section, we solve the Boltzmann equations
and obtain the spin signal defined in eq.~(\ref{eq:def-Rs})
by evaluating $V_{\rm p}$ and $V_{\rm ap}$.
Note that $V_{\rm p}$ ($V_{\rm ap}$) is the open-circuit voltage induced across
the detection junction when the magnetizations of the injector and detector
are in the parallel (antiparallel) alignment.
We determine $V_{\rm p}$ and $V_{\rm ap}$
by the condition of $I_{\rm det} = 0$.
We assume that the magnitude of $V_{\rm p}$ and $V_{\rm ap}$
is much smaller than $\Delta/e$.
However, we do not assume $V_{\rm inj} \ll \Delta/e$.
We focus on the case where the injection current is so small that
injected quasiparticles are populated only near the gap edge
(i.e., $|\epsilon| \approx \Delta$).
In this case, $f_{{\rm T}+}$ and $f_{{\rm T}-}$ quickly relaxes
because the conversion time becomes very short near
the gap edge.~\cite{rf:tinkham,rf:schmid,rf:takane3,rf:takane4}
Therefore, we ignore $f_{{\rm T}+}$ and $f_{{\rm T}-}$
in the following argument.
Furthermore, the smallness of the injection current also allows us to ignore
$f_{{\rm L}+}(x_{\rm inj},\epsilon)$ and $f_{{\rm L}-}(x_{\rm inj},\epsilon)$
in the injection terms given in eqs.~(\ref{eq:PL+}) and (\ref{eq:PL-}).

We first assume that magnetic impurities are absent
(i.e., $\tau_{\rm m}^{-1} = 0$)
and define the spin-flip scattering time $\tau_{\rm sf}$ as
\begin{align}
  \frac{1}{\tau_{\rm sf}} = \frac{4}{3\tau_{\rm so}} .
\end{align}
In this case, the spectral functions for $|\epsilon| > \Delta$ are simply
given by
\begin{align}
      \label{eq:N1-0}
   N_{1}(\epsilon) & = \frac{|\epsilon|}
                            {\sqrt{\epsilon^{2}-\Delta^{2}}} ,
     \\
      \label{eq:R2-0}
   R_{2}(\epsilon) & = \frac{{\rm sign} (\epsilon) \Delta}
                            {\sqrt{\epsilon^{2}-\Delta^{2}}} ,
\end{align}
and $N_{2}(\epsilon) = R_{1}(\epsilon) = 0$.
This indicates that $N_{1}^{2}(\epsilon)-R_{2}^{2}(\epsilon) = 1$.
We first obtain $f_{{\rm L}+}(x_{\rm det},\epsilon)$
by solving eq.~(\ref{eq:fL+}).
Note that $f_{{\rm L}+}(x,\epsilon)$ decays exponentially as a function of
$|x-x_{\rm inj}|$ and this decay is characterized by the spin-diffusion length
given by $\lambda_{\rm sf} = \sqrt{D_{\rm S}\tau_{\rm sf}}$.
We obtain
\begin{align}
     \label{eq:f_+det}
 f_{{\rm L}+}(x_{\rm det},\epsilon)
  = \frac{P_{\rm spin}\lambda_{\rm sf}}
         {8e^{2}N_{\rm S}A_{\rm S}R_{\rm inj}D_{\rm S}}
    \Sigma_{+}(\epsilon,V_{\rm inj}) {\rm e}^{-\frac{d}{\lambda_{\rm sf}}}
\end{align}
with
\begin{align}
   \Sigma_{+}(\epsilon,V_{\rm inj})
     = N_{1}(\epsilon) \bigl(  f_{0} \left(\epsilon-eV_{\rm inj}\right)
                             - f_{0} \left(\epsilon+eV_{\rm inj}\right)
                       \bigr) .
\end{align}
Next, we obtain $f_{{\rm L}-}(x_{\rm det},\epsilon)$ which is necessary
to obtain $f_{{\rm F}+}(y_{\rm det},\epsilon)$.
A special care must be paid in solving eq.~(\ref{eq:fL-})
since no relaxation process is included in this equation.
The relaxation of $f_{{\rm L}-}$ is mainly caused by
the phonon-mediated recombination process, which is described by adding
the following nonlinear term~\cite{rf:takane2}
\begin{align}
  I_{{\rm L}-}(x,\epsilon)
  & = - 4 \int {\rm d} \epsilon' \sigma_{\rm S}(\epsilon,\epsilon')
     \nonumber \\
  & \hspace{-5mm}
    \times
         \left(  N_{1}(\epsilon)N_{1}(\epsilon')
               - R_{2}(\epsilon)R_{2}(\epsilon') \right)
         f_{{\rm L}-}(x,\epsilon)f_{{\rm L}-}(x,\epsilon')
\end{align}
to eq.~(\ref{eq:fL-}).
Here, $\sigma_{\rm S}(\epsilon,\epsilon')$ is identical to
$\sigma_{\rm F}(\epsilon,\epsilon')$ in eq.~(\ref{eq:sigma_F})
if $\alpha_{\rm F}$ is replaced by $\alpha_{\rm S}$.
From this expression, we observe that the corresponding decay length
$L_{\rm c}$ becomes very long when the injection current is small
and therefore $|f_{{\rm L}-}(x,\epsilon)| \ll 1$.
We thus assume that $L_{\rm c}$ is longer than, or at least of the order of,
the length of the superconducting wire, and adopt the boundary condition that
$f_{{\rm L}-}$ vanishes at each end of the superconducting wire.
We further assume that the distance between the injection junction and
each end of the superconductor is nearly equal to $L$, and $L \gg d$.
Under this assumption, we approximately obtain
\begin{align}
    \label{eq:f_-det}
 f_{{\rm L}-}(x_{\rm det},\epsilon)
  = \frac{L}{8e^{2}N_{\rm S}A_{\rm S}R_{\rm inj}D_{\rm S}}
    \Sigma_{-}(\epsilon,V_{\rm inj})
\end{align}
with
\begin{align}
   \Sigma_{-}(\epsilon,V_{\rm inj})
  &   = N_{1}(\epsilon) \bigl(  f_{0} \left(\epsilon-eV_{\rm inj}\right)
                             + f_{0} \left(\epsilon+eV_{\rm inj}\right)
     \nonumber \\
  & \hspace{30mm}
                             - 2 f_{0} \left(\epsilon\right)
                       \bigr) .
\end{align}
If $L \gg L_{\rm c}$, we must replace $L$ in eq.~(\ref{eq:f_-det})
with $L_{\rm c}$.
Finally, we obtain $f_{{\rm F}+}(y_{\rm det},\epsilon)$.
It should be emphasized that $I_{\rm F}$ containing $f_{{\rm F}+}$ becomes
relevant in the low temperature regime where the ordinary contribution
$I_{\rm q}$ is exponentially suppressed due to the opening of the energy gap.
In this regime, the term with $f_{0}(\epsilon)-f_{0}(\epsilon-eV_{\rm det})$
in $P_{{\rm F}+}$ can be neglected.
Furthermore, since $f_{{\rm T}+}$ can be ignored,
the dominant contribution to $P_{{\rm F}+}$
arises from the term with $f_{{\rm L}-}$.
This indicates that nonequilibrium quasiparticles are created
by energy imbalance in the superconductor.
We thus approximate $P_{{\rm F}+}$ as
\begin{align}
     P_{{\rm F}+}(y,\epsilon)
   = \frac{\delta(y-y_{\rm det})N_{1}(\epsilon)}
          {2e^{2}N_{\rm F}A_{\rm F}R_{\rm det}}
     f_{{\rm L}-}(x_{\rm det},\epsilon) .
\end{align}
Solving eq.~(\ref{eq:fF+}), we obtain
\begin{align}
     f_{{\rm F}+}(y_{\rm det},\epsilon)
   = \frac{\lambda_{\rm e}(\epsilon-eV_{\rm det})}
          {4e^{2}N_{\rm F}A_{\rm F}R_{\rm det}D_{\rm F}}
     N_{1}(\epsilon) f_{{\rm L}-}(x_{\rm det},\epsilon) ,
\end{align}
where the energy relaxation length $\lambda_{\rm e}$ is given by
$\lambda_{\rm e}(\epsilon) = \sqrt{D_{\rm F}\tau_{\rm e}(\epsilon)}$.
Combining this and eq.~(\ref{eq:f_-det}) and noting that quasiparticles are
populated near the gap edge (i.e., $|\epsilon| \approx \Delta$),
we approximately obtain
\begin{align}
       \label{eq:f(+)+f(-)}
 & f_{{\rm F}+}(y_{\rm det},\epsilon) + f_{{\rm F}+}(y_{\rm det},-\epsilon)
       \nonumber \\
 & \hspace{5mm}
   = \frac{L}{8e^{2}N_{\rm S}A_{\rm S}R_{\rm inj}D_{\rm S}}
     \frac{\lambda_{\rm e}(\Delta)}
          {4e^{2}N_{\rm F}A_{\rm F}R_{\rm det}D_{\rm F}}
       \nonumber \\
 & \hspace{10mm} \times
     N_{1}(\epsilon) \Sigma_{-}(\epsilon,V_{\rm inj})
     \frac{3eV_{\rm det}}{\Delta} .
\end{align}
From eqs.~(\ref{eq:I_F_mod}) and (\ref{eq:f(+)+f(-)}), we observe that
$I_{\rm F}= 0$ at $V_{\rm det} = 0$.
This reflects the fact that the quasiparticle distribution $f_{{\rm F}+}$
created by the tunneling of energy-imbalanced quasiparticles
can contribute to the tunneling current only when
the energy relaxation time for $f_{{\rm F}+}(y,\epsilon)$ is different
from that for $f_{{\rm F}+}(y,-\epsilon)$.~\cite{rf:takane2}
That is, the energy relaxation process is  essential
in obtaining a nonzero $I_{\rm F}$.

We obtain $I_{\rm F}$ and $I_{\rm S}$ by substituting the resulting
quasiparticle distributions into eqs.~(\ref{eq:I_S}) and (\ref{eq:I_F_mod}).
The three terms are given as follows:
\begin{align}
  I_{\rm q} & = \chi(T) \frac{V_{\rm det}}{R_{\rm det}} ,
        \\
  I_{\rm F} & = \frac{3R_{\rm S}R_{\rm F}}{8R_{\rm inj}R_{\rm det}}
                \frac{L\lambda_{\rm e}(\Delta)}{\lambda_{\rm sf}^{2}}
                J_{3}(V_{\rm inj},T) \frac{V_{\rm det}}{R_{\rm det}} ,
        \\
  I_{\rm S} & = \eta
                \frac{R_{\rm S}}{2R_{\rm inj}}
                P_{\rm spin}^{2}{\rm e}^{-\frac{d}{\lambda_{\rm sf}}}
                J_{2}(V_{\rm inj},T) \frac{\Delta}{eR_{\rm det}} ,
\end{align}
where $\eta = 1 (-1)$ for the parallel (antiparallel) alignment and
\begin{align}
  \chi(T) & = \int_{0}^{\infty}{\rm d}\epsilon N_{1}(\epsilon)
              \left(-2\frac{\partial f_{0}(\epsilon)}{\partial \epsilon}
              \right) ,
       \\
  J_{3}(V,T)
          & = \frac{1}{\Delta}
              \int_{0}^{\infty}{\rm d}\epsilon N_{1}^{3}(\epsilon)
              \bigl(  f_{0} \left(\epsilon-eV\right)
                    + f_{0} \left(\epsilon+eV\right)
       \nonumber \\
 & \hspace{40mm}
                    - 2 f_{0} \left(\epsilon\right)
              \bigr) ,
       \\
  J_{2}(V,T)
          & = \frac{1}{\Delta}
              \int_{0}^{\infty}{\rm d}\epsilon N_{1}^{2}(\epsilon)
              \bigl(  f_{0} \left(\epsilon-eV\right)
                    - f_{0} \left(\epsilon+eV\right)
              \bigr) .
\end{align}
The resistances $R_{\rm S}$ and $R_{\rm F}$ are defined by
$R_{\rm S} \equiv \lambda_{\rm sf}\rho_{\rm S}/A_{\rm S}$ and
$R_{\rm F} \equiv \lambda_{\rm sf}\rho_{\rm F}/A_{\rm F}$
with the resistivities $\rho_{\rm S} = (2e^{2}N_{\rm S}D_{\rm S})^{-1}$
and $\rho_{\rm F} = (2e^{2}N_{\rm F}D_{\rm F})^{-1}$.
It should be noted that $J_{3}(V,T)$ and $J_{2}(V,T)$ diverge
if eq.~(\ref{eq:N1-0}) is adopted as the expression of $N_{1}(\epsilon)$.
This unphysical divergence does not arise if we adopt a more realistic
expression of $N_{1}(\epsilon)$, which does not diverges at the gap edge.
Indeed, the divergence of $N_{1}(\epsilon)$ is actually removed
if we take account of gap anisotropy, inelastic electron scattering
or magnetic impurity scattering.
We obtain $V_{\rm p}$ and $V_{\rm ap}$ by solving
$I_{\rm det}(V_{\rm p}) = 0$ for $\eta = 1$ and
$I_{\rm det}(V_{\rm ap}) = 0$ for $\eta = - 1$, respectively.
Substituting the resulting expressions and eq.~(\ref{eq:I_inj})
into eq.~(\ref{eq:def-Rs}), we finally obtain
\begin{align}
    \label{eq:R_spin-result}
 R_{\rm spin} = \gamma
                P_{\rm spin}^{2}R_{\rm S}{\rm e}^{-\frac{d}{\lambda_{\rm sf}}}
\end{align}
with
\begin{align}
    \label{eq:gamma-result}
 \gamma = \frac{J_{2}(V_{\rm inj},T)}
               {J_{1}(V_{\rm inj},T)
                \left( \chi(T)+\frac{3R_{\rm S}R_{\rm F}}
                                    {8R_{\rm inj}R_{\rm det}}
                               \frac{L\lambda_{\rm e}(\Delta)}
                                    {\lambda_{\rm sf}^{2}} J_{3}(V_{\rm inj},T)
                \right)} .
\end{align}
Note that $\gamma$ represents the renormalization of the spin signal
induced by the transition to the superconducting state,
and $\gamma = 1$ corresponds to the normal state.

In the remaining of this section, we briefly consider the influence of
magnetic impurities.
We redefine $\tau_{\rm sf}$ as
\begin{align}
 \frac{1}{\tau_{\rm sf}}
   = \frac{4}{3\tau_{\rm so}} + \frac{4}{3\tau_{\rm m}} ,
\end{align}
and introduce the parameter~\cite{rf:poli}
\begin{align}
 \beta
   = \frac{\tau_{\rm so}-\tau_{\rm m}}{\tau_{\rm so}+\tau_{\rm m}}
\end{align}
which characterizes the relative strength of spin-orbit scattering
and magnetic impurity scattering.
Here, $\tau_{\rm sf}$ should be regarded as the spin-flip scattering time
in the normal state.
We observe that $\beta = - 1$ in the absence of magnetic impurities
and $\beta = 1$ when spin-orbit scattering does not occur.
If $\beta \neq - 1$, we must solve eq.~(\ref{eq:usadel})
to obtain the spectral functions.
Strictly speaking, eqs.~(\ref{eq:N1-0}) and (\ref{eq:R2-0}) are not justified
in the presence of magnetic impurities and the relation
$N_{1}^{2}(\epsilon)-R_{2}^{2}(\epsilon) = 1$ no longer holds exactly.
Consequently, $f_{{\rm L}+}(x_{\rm det},\epsilon)$ and
$f_{{\rm L}-}(x_{\rm det},\epsilon)$ are modified as
\begin{align}
     \label{eq:f_+det-mag}
 f_{{\rm L}+}(x_{\rm det},\epsilon)
 & = \frac{P_{\rm spin}\alpha(\epsilon)\lambda_{\rm sf}}
          {8e^{2}N_{\rm S}A_{\rm S}R_{\rm inj}D_{\rm S}}
     \frac{\Sigma_{+}(\epsilon,V_{\rm inj})}
          {(N_{1}^{2}(\epsilon)-R_{2}^{2}(\epsilon))}
       \nonumber \\
 & \hspace{30mm} \times
     {\rm e}^{-\frac{d}{\alpha(\epsilon)\lambda_{\rm sf}}} ,
      \\
    \label{eq:f_-det-mag}
 f_{{\rm L}-}(x_{\rm det},\epsilon)
 & = \frac{L}{8e^{2}N_{\rm S}A_{\rm S}R_{\rm inj}D_{\rm S}}
     \frac{\Sigma_{-}(\epsilon,V_{\rm inj})}
          {(N_{1}^{2}(\epsilon)-R_{2}^{2}(\epsilon))} ,
\end{align}
where
\begin{align}
  \alpha(\epsilon)
 = \sqrt{\frac{N_{1}^{2}(\epsilon)-R_{2}^{2}(\epsilon)}
              {N_{1}^{2}(\epsilon)+\beta R_{2}^{2}(\epsilon)}} .
\end{align}
The parameter $\alpha(\epsilon)$ represents the renormalization of
the spin-flip scattering time
on transition to the superconducting state.~\cite{rf:morten2}
Using eqs.~(\ref{eq:f_+det-mag}) and (\ref{eq:f_-det-mag}),
we can show that eqs.~(\ref{eq:R_spin-result}) and (\ref{eq:gamma-result})
are applicable to this case
if $J_{3}$ and $J_{2}$ are replaced by the following expressions,
\begin{align}
  J_{3}(V,T)
          & = \frac{1}{\Delta}
              \int_{0}^{\infty}{\rm d}\epsilon
              \frac{N_{1}^{3}(\epsilon)}
                   {N_{1}^{2}(\epsilon)-R_{2}^{2}(\epsilon)}
              \bigl(  f_{0} \left(\epsilon-eV\right)
       \nonumber \\
 & \hspace{20mm}
                    + f_{0} \left(\epsilon+eV\right)
                    - 2 f_{0} \left(\epsilon\right)
              \bigr) ,
       \\
  J_{2}(V,T)
          & = \frac{1}{\Delta}
              \int_{0}^{\infty}{\rm d}\epsilon
              \frac{\alpha(\epsilon)N_{1}^{2}(\epsilon)}
                   {N_{1}^{2}(\epsilon)-R_{2}^{2}(\epsilon)}
              {\rm e}^{-(\alpha(\epsilon)^{-1}-1)\frac{d}{\lambda_{\rm sf}}}
       \nonumber \\
 & \hspace{10mm} \times
              \bigl(  f_{0} \left(\epsilon-eV\right)
                    - f_{0} \left(\epsilon+eV\right)
              \bigr) .
\end{align}

We here comment on the expression of the spin signal
presented by Poli \textit{et al}.~\cite{rf:poli}
We note that they ignore the influence of nonequilibrium quasiparticles
in the ferromagnetic metal for detection
and therefore the corresponding term is lacking.
This is the significant difference between their expression and ours.
In addition, they assume $V_{\rm inj} \ll \Delta/e$.
Finally, we point out that $\alpha(\epsilon)$-dependence is slightly
different between them.
Indeed, if the factor $2\alpha + N(E)R_{N}/R_{I}$ in eq.~(4) of
ref.~\citen{rf:poli} is replaced by $2\alpha^{-1}$,
their expression becomes nearly identical to ours
in the case of $J_{3}(V_{\rm inj},T) = 0$ and $V_{\rm inj} \ll \Delta/e$.
The reason for this difference is not clear.

\section{Discussion}

Let us consider the temperature dependence of
the renormalization factor $\gamma$
under the condition that the injection current $I_{\rm inj}$ is kept constant.
We adjust $V_{\rm inj}$ to supply a constant injection current.
This means that $V_{\rm inj}$ is determined as a function of $T$
for a given $I_{\rm inj}$, so we rewrite $J_{i}(V_{\rm inj},T)$ as
$J_{i}(I_{\rm inj},T)$ ($i = 1,2,3$).
It should be noted here that even though $I_{\rm inj}$ is very small,
$V_{\rm inj}$ approaches to $\Delta/e$ as $T \to 0$.

We focus on the low temperature regime
where the $T$-dependence of $\Delta$ can be neglected.
In this regime, $\chi(T)$ behaves as
\begin{align}
    \label{eq:chai-LT}
  \chi(T) = \sqrt{\frac{2\pi\Delta}{T}} {\rm e}^{-\frac{\Delta}{T}} .
\end{align}
When $T$ is not very low and $\chi(T)$ is much greater than the term with
$J_{3}(I_{\rm inj},T)$ in the denominator of eq.~(\ref{eq:gamma-result}),
the renormalization factor is reduced to
\begin{align}
    \label{eq:gamma-result-HT}
 \gamma = \frac{J_{2}(I_{\rm inj},T)}
               {J_{1}(I_{\rm inj},T) \chi(T)} .
\end{align}
Because the $T$-dependence of $J_{2}(I_{\rm inj},T)/J_{1}(I_{\rm inj},T)$
is weak, we obtain $\gamma \propto \chi(T)^{-1}$.
This indicates that $\gamma$ behaves as $\gamma \propto {\rm e}^{\Delta/T}$.
However, because $\chi(T)$ is exponentially suppressed with decreasing $T$,
the term with $J_{3}(I_{\rm inj},T)$ eventually dominates $\chi(T)$ below
a crossover temperature $T_{\rm cross}$.
Below $T_{\rm cross}$, we can ignore $\chi(T)$ in
eq.~(\ref{eq:gamma-result}) and the renormalization factor is reduced to
\begin{align}
    \label{eq:gamma-result-LT}
 \gamma = \frac{8R_{\rm inj}R_{\rm det}}{3R_{\rm S}R_{\rm F}}
          \frac{\lambda_{\rm sf}^{2}}{L\lambda_{\rm e}(\Delta)}
          \frac{J_{2}(I_{\rm inj},T)J_{3}(I_{\rm inj},T)}
               {J_{1}(I_{\rm inj},T)} .
\end{align}
The crossover temperature is determined by
\begin{align}
    \label{eq:T_cross}
   \chi(T_{\rm cross})
           = \frac{3R_{\rm S}R_{\rm F}}
                  {8R_{\rm inj}R_{\rm det}}
             \frac{L\lambda_{\rm e}(\Delta)}
                  {\lambda_{\rm sf}^{2}} J_{3}(I_{\rm inj},T_{\rm cross}) .
\end{align}
As $T$ is lowered below $T_{\rm cross}$,
the injection voltage $V_{\rm inj}$ approaches to $\Delta/e$.
In this situation, the $T$-dependence of $J_{3}(I_{\rm inj},T)$ becomes weak.
Furthermore, we can neglect the weak $T$-dependence
of $J_{2}(I_{\rm inj},T)/J_{1}(I_{\rm inj},T)$.
Thus, we conclude that below $T_{\rm cross}$,
the renormalization factor $\gamma$ rapidly converges
to the value given by $\gamma_{0} \equiv \lim_{T \to 0} \gamma$.
We can obtain $\gamma_{0}$ from eq.~(\ref{eq:gamma-result-LT}) with $T = 0$.

From the above argument, we observe the qualitative behavior
of $R_{\rm spin}$ as follows.
In the regime of $T \gg T_{\rm cross}$, the spin signal exponentially increases
with decreasing $T$ as $R_{\rm spin} \propto {\rm e}^{\Delta/T}$.
Below $T_{\rm cross}$, however, the spin signal converges as
$R_{\rm spin} \to \gamma_{0}
P_{\rm spin}^{2}R_{\rm S}{\rm e}^{-d/\lambda_{\rm sf}}$.
We here point out that the behavior of $R_{\rm spin}$ in the regime of
$T \gg T_{\rm cross}$ is qualitatively equivalent to the previous result,
eq.~(\ref{eq:TM}), reported by Takahashi and Maekawa.~\cite{rf:takahashi}
However, our argument indicates that
the exponential increase of $R_{\rm spin}$ should not be attributed to
the increase of spin resistivity.~\cite{rf:takahashi}
We simply understand that $R_{\rm spin}$ increases reflecting
the suppression of thermally excited quasiparticles in the detection junction.

Let us consider the experimental result reported by Poli
\textit{et al}.~\cite{rf:poli} on the basis of our theoretical framework.
Particularly, we focus on the convergence of $R_{\rm spin}$
observed at low temperatures.
They employed the device consisting of a superconducting wire of Al
and ferromagnetic metal wires of Co.
Since it has been believed that spin-flip scattering in Al is
mainly caused by spin-orbit interaction,
we assume that magnetic impurity scattering is much less relevant than
spin-orbit scattering and set $\tau_{\rm m}^{-1} = 0$.
We estimate the limiting value $\gamma_{0}$ of the renormalization factor
from eq.~(\ref{eq:gamma-result-LT}) with $T = 0$
and compare it with their experimental value.
Following refs.~\citen{rf:urech} and \citen{rf:poli},
we employ the parameters:
$I_{\rm inj} = 1 \ {\rm nA}$,
$\lambda_{\rm sf} = 1 \ \mu {\rm m}$,
$\Delta = 200 \ \mu {\rm eV}$,
$R_{\rm inj} = R_{\rm det} = 100 \ {\rm k}\Omega$,
$\rho_{\rm S} = 10 \ \mu \Omega {\rm cm}$,
$A_{\rm S} = 10 \times 150 \ {\rm nm}^{2}$,
$A_{\rm F} = 50 \times 130 \ {\rm nm}^{2}$.
For the other parameters, we assume
$L = 10 \ \mu {\rm m}$,
$D_{\rm F} = 3 \times 10^{-3} \ {\rm m^{2}s^{-1}}$,
$\rho_{\rm F} = 14 \ \mu \Omega {\rm cm}$,
$\alpha_{\rm F} = 9 \times 10^{3} \ {\rm eV}^{-2}$.
The value of $\alpha_{\rm F}$ is estimated by using the
relation~\cite{rf:tinkham} $\alpha_{\rm F} \sim 2/\tau_{\rm D}T_{\rm D}^{3}$
with $T_{\rm D} = 385 \ {\rm K}$ and
$\tau_{\rm D} = 0.4 \times 10^{-14} \ {\rm s}$,
where $T_{\rm D}$ and $\tau_{\rm D}$ are the Debye temperature and
the phonon scattering time at $T_{\rm D}$, respectively.
From these parameters, we obtain
$\lambda_{\rm e}(\Delta) = 9 \ \mu {\rm m}$,
$R_{\rm S} = 67 \ \Omega$ and $R_{\rm F} = 22 \ \Omega$.
The integral $J_{1}(I_{\rm inj},T)$ does not depend on $T$
and is obtained from eq.~(\ref{eq:I_inj}) as
$J_{1}(I_{\rm inj},T) = eR_{\rm inj}I_{\rm inj}/\Delta = 0.5$.
We finally consider $J_{2}(I_{\rm inj},T)$ and $J_{3}(I_{\rm inj},T)$
in the limit of $T \to 0$.
The evaluation of these integrals is not simple,
so we roughly approximate them as
$J_{2}(I_{\rm inj},0) = J_{3}(I_{\rm inj},0) = J_{1}(I_{\rm inj},T)$.
Substituting these parameters into eq.~(\ref{eq:gamma-result-LT}),
we approximately obtain $\gamma_{0} \sim 10^{5}$.
This indicates that $R_{\rm spin}$ below $T_{\rm cross}$ is
a factor of $10^{5}$ larger than that in the normal state.
This is consistent with the experimental result which indicates
the enhancement of $4$ or $5$ orders of magnitude.
We estimate the crossover temperature by solving eq.~(\ref{eq:T_cross}) with
eq.~(\ref{eq:chai-LT}) and obtain $T_{\rm cross} \sim 0.1 \ {\rm K}$.
This is also consistent with the experimental value of
$T_{\rm cross} \sim 0.16 \ {\rm K}$.

We have shown that nonequilibrium quasiparticles
with $|\epsilon| \approx \Delta$ are created in the ferromagnetic metal
for detection by the tunneling of energy-imbalanced quasiparticles,
and that these quasiparticles contribute to $I_{\rm det}$ in combination with
the energy relaxation process due to phonon scattering.
It should be noted that the energy relaxation of quasiparticles
excites phonons near the detection junction, leading to
the increase of effective temperature $T_{\rm eff}$ for quasiparticles.
If $T_{\rm eff}$ becomes greater than $T_{\rm cross}$,
the convergence of the spin signal is determined by this heating effect
instead of the convergence mechanism which we discussed above.
The separation of these two mechanisms is a future problem for experiments.

In addition to the heating effect, we have ignored charge imbalance.
If the injection current is not small, we must consider its influences.
Charge imbalance provides a nearly constant contribution $I_{\rm Q}$
to $I_{\rm S}$ regardless of the alignment of magnetizations.
Since $I_{\rm Q}$ must be cancelled by $I_{\rm q}$ and $I_{\rm F}$
to ensure $I_{\rm det} = 0$, we expect that
both $V_{\rm p}$ and $V_{\rm ap}$ increases with increasing $I_{\rm Q}$.
However, if $I_{\rm Q}$ is sufficiently small,
the increase of $V_{\rm p}$ is equivalent to that of $V_{\rm ap}$
because both $I_{\rm q}(V_{\rm det})$ and $I_{\rm F}(V_{\rm det})$ linearly
depends on $V_{\rm det}$ when $|V_{\rm det}| \ll \Delta/e$.
Therefore, we expect that no qualitative change of the spin signal
appears as long as charge imbalance is not very large.

In summary, we have studied the transport of spin-polarized nonequilibrium
quasiparticles in a superconducting wire connected by tunnel junctions
to two ferromagnetic metal wires,
each of which serves as a spin injector or detector.
We have presented a basic formalism to determine spin-polarized quasiparticle
distributions in this system, and obtained an analytical expression
for the nonlocal spin signal.
We have taken account of nonequilibrium quasiparticles in the ferromagnetic
metal for detection, which are created by the tunneling of energy-imbalanced
quasiparticles in the superconductor.
We have shown that they induce the convergence of
the spin signal at low temperatures.

\end{document}